\documentclass{article}
\usepackage{dcase2024,amsmath,graphicx,url,times,booktabs, tabularx, comment, hyperref}
\usepackage{caption}
\usepackage{subcaption}

\title{Synthetic training set generation using text-to-audio models for environmental sound classification}

\name{Francesca Ronchini, %$^{1}$,
       Luca Comanducci, %$^{1}$,
       Fabio Antonacci%$^{1}$, 
       }
 \address{Dipartimento di Elettronica, Informazione e Bioingegneria, Politecnico di Milano, Italy, \\\{francesca.ronchini, luca.comanducci, fabio.antonacci\}@polimi.it \\          
  }

\begin{document}

\ninept
\maketitle

\begin{sloppy}

\begin{abstract}
In recent years, text-to-audio models have revolutionized the field of automatic audio generation. This paper investigates their application in generating synthetic datasets for training data-driven models. Specifically, this study analyzes the performance of two environmental sound classification systems trained with data generated from text-to-audio models. We considered three scenarios: a) augmenting the training dataset with data generated by text-to-audio models; b) using a mixed training dataset combining real and synthetic text-driven generated data; and c) using a training dataset composed entirely of synthetic audio. In all cases, the performance of the classification models was tested on real data. Results indicate that text-to-audio models are effective for dataset augmentation, with consistent performance when replacing a subset of the recorded dataset. However, the performance of the audio recognition models drops when relying entirely on generated audio.
\end{abstract}


\section{Introduction}
In the past few years, Text-To-Audio (TTA) models have become the new state-of-the-art for what concerns machine learning-based sound synthesis. TTA models are deep learning generative systems designed to generate audio samples based on textual descriptions, commonly referred to as prompts, which are given as input to the models.
Several TTA models have been proposed to generate high-quality, realistic audio samples. Pioneering models include AudioGen~\cite{kreuk2022audiogen}, an auto-regressive generative model, and AudioLDM~\cite{liu2023audioldm}, based on a latent diffusion model~\cite{rombach2022high}. AudioLDM2~\cite{liu2023audioldm2}, a more sophisticated version of AudioLDM, has been recently proposed. Other TTA systems include Tango~\cite{ghosal2023text}, Make-an-Audio~\cite{huang2023make}, %, huang2023make2%Auffusion~\cite{xue2024auffusion},
and Audiobox~\cite{vyas2023audiobox}. 

Thanks to the high-quality generated audio and ease of use, TTA models have been applied to a wide variety of diverse domains such as augmented and virtual reality~\cite{nordahl2011sound}, foley sound generation~\cite{chung2024t}, among others. The versatility of these models makes them potentially applicable to the task of synthetic dataset generation or data augmentation for deep learning models, particularly in cases where data collection is challenging due to privacy concerns or limited data availability.
In fact, one limitation of data-driven approaches is the need for large amounts of labeled training data to reach good performances. Unfortunately, dataset acquisition and labeling are time-consuming and biases-prone procedures~\cite{ronchini2021impact}. Several studies in the field of sound recognition have shown that augmenting the original dataset with synthetic data during the training phase improves system generalization and enhances performances~\cite{salamon2017scaper, ronchini2022benchmark, gontier2021polyphonic, damiano, ibrahim2024towards}. The synthetic data considered in previous works were generated using signal-processing-based or audio-mixing tools for synthesizing soundscapes, such as Scaper~\cite{salamon2017scaper} or Pyroadacoustics~\cite{damiano2022pyroadacoustics}. These techniques require the manual tuning of different parameters of the sound generation procedure~\cite{salamon2017scaper, damiano2022pyroadacoustics}, potentially making the process even more cumbersome and error-prone. The introduction of TTA models could be beneficial as they have the potential to overcome these limitations by allowing the generation of the desired audio content through natural language. However, the literature related to the use of TTA for dataset generation is still limited. In~\cite{kroher2023can}, Kroher et al. trained a music genre classifier on a fully artificial music dataset generated with MusicGen~\cite{copet2024simple}, a text-to-music generation model. The study focused on 5 music genres and the results show that the classifier effectively generalized features learned from artificial data to real music recordings. In~\cite{zhang2023first}, the authors fine-tuned AudioLDM to generate both normal and anomalous sounds, which were included in the training dataset for the anomalous sound detection task. The results indicate that generative sounds are promising to achieve performances comparable to state-of-the-art models.  

Motivated by these positive preliminary findings~\cite{kroher2023can, zhang2023first}, this paper investigates how to leverage 
TTA models in the field of Environmental Sound Classification (ESC)~\cite{bansal2022environmental}. ESC refers to the task of classifying environmental sounds that can be presented in an audio clip. % given as input to an ESC system.
In our opinion, ESC is an ideal application area to investigate the possibility of including TTA-generated synthetic data for two reasons: TTA models can generate all the sound types present in most ESC datasets; ESC can be considered between the simplest scenarios among the ones considered by the DCASE community. Therefore, it is naturally the first one to address before tackling more complex tasks. 

Concurrently to our work, a similar research study addressed the problem applied to speech modeling and audio recognition~\cite{feng2024can}; here we specifically focus on the ESC task. We consider two state-of-the-art deep learning models for ESC and analyze how their performances vary when TTA-generated data are included as part of the training dataset according to different methodologies: 1) using TTA to perform data augmentation; 2) using TTA data as the sole source of training data; 3) mixing TTA-generated and real data. Audio samples and the code used for this study are available on GitHub\footnote{\href{https://ronfrancesca.github.io/Text-to-Audio-ESC/}{https://ronfrancesca.github.io/Text-to-Audio-ESC/}}.

\section{Experimental procedure}
\label{sec:task}
In this section, we briefly introduce the TTA models selected for the dataset generation, the prompt strategies for generating it, and the dataset generation process. Sec.~\ref{subsec:models} briefly introduces the used ESC model architectures. 

\subsection{Text-to-Audio models}
\label{subsec:TTA}
We selected two pre-trained models for generating the dataset: AudioGen~\cite{kreuk2022audiogen} and AudioLDM2~\cite{liu2023audioldm2}.

AudioGEN is an auto-regressive model that learns a discrete representation of raw audio through an auto-encoding procedure. It then generates audio using a transformer model applied to the learned representation, conditioned on textual features~\cite{kreuk2022audiogen}.

AudioLDM2 is a continuous latent-diffusion model conditioned via CLAP~\cite{wu2023large}, which removes the need for paired audio-text data during the training process. 

\subsection{Synthetic dataset generation process}
\label{subsec:datagen}
UrbanSound8K (US8K)~\cite{salamon2014dataset} is the dataset selected for this study. Along with ESC-10 and ESC-50~\cite{piczak2015esc}, they serve as the primary datasets used as benchmarks for ESC tasks. The size of US8K makes it more appropriate for training deep learning models compared to ESC-10 and ESC-50, mainly used for evaluation.
US8K contains $8732$ labeled sounds of $4~\mathrm{s}$ maximum duration of urban sounds from $10$ classes. The dataset is divided into $10$ folds, used for leave-one-out cross-validation at evaluation time. Fig.~\ref{fig:class_distribution} reports the sound classes and their distribution in folders. We generated four versions of the US8K dataset: two with AudioLDM2 and two with AudioGen. For each of them, we first generated the total amount of data and then randomly divided it into $10$ folds, following the same distribution of US8K. 
\begin{figure}[t!]
\centering
\includegraphics[width=.8\columnwidth]{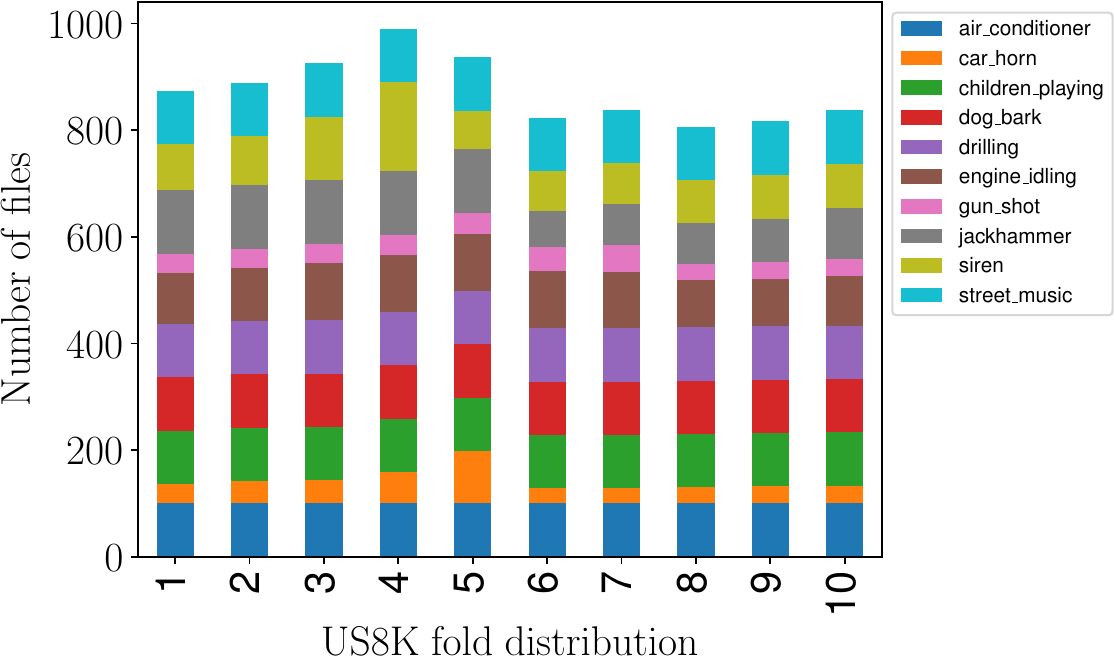}
\caption{US8K dataset classes distribution per each fold. Colors represent the different sound classes, specified in the legend.}
\label{fig:class_distribution}
\end{figure}

\subsection{Prompt templates}
\label{subseb:prompts}
We generated the dataset using two prompt strategies. Both strategies employ a single-instruction sentence containing a sound generation instruction in an urban context, specifying the desired audio class. We explored various templates and informally listened to the generated audio to determine which prompt was the most effective.
The first strategy's template is:~\textit{``A clear sound of a $<$class\_to\_generate$>$ in an urban context."} The template and the adjective choice are based on suggestions from the AudioLDM2 authors' guidelines\footnote{\label{note1}{\href{https://huggingface.co/docs/diffusers/main/en/api/pipelines/audioldm2}{https://huggingface.co/docs/diffusers/main/en/api/pipelines/audioldm2}}} and studies related to prompt tuning for sound classification~\cite{elizalde2023clap}. %We explored various templates and listened to the generated audio to determine which prompt was the most effective.
Notably, the use of the adjective \textit{clear} is supported by frequency counts of training data and by its use in other studies~\cite{deshmukh2024pam}. The second strategy uses ChatGPT 3.5 Large Language Model (LLM), which is asked to generate a single sentence to be used as input for a TTA model. Different studies have shown that using an LLM provides diverse and contextually rich prompts~\cite{he2022synthetic, mei2023wavcaps}. The prompt template suggested by the LLM was: \textit{``Generate a realistic audio representation of the sound of a $<$class\_to\_generate$>$ in an urban environment"}. For AudioLDM2 we also used \textit{``Low quality"} as a negative prompt, following the authors' guidelines and implementations in other domains~\cite{voetman2023using}.
Depending on the sound class, the templates were adapted to include repetitive sounds for consistency with the study's padding strategy (e.g., dog bark or car horn) or to better specify a sound that might confuse its generation (e.g., siren). The same templates were used for AudioGen to ensure consistency. However, AudioGen does not involve the use of a negative prompt. 
%We would like to specify that 
We are conscious of the fact that handcrafted prompts proposed to generate the data could be a limitation of the current study~\cite{deshmukh2024domain}. Alternative prompt strategies will be considered in future works. 

\begin{table}[t!]
\centering
  \caption{Data augmentation comparison between signal-processing-based and TTA-based strategies.}
  \footnotesize
  \begin{tabular}{l|c|c}
  \toprule
  Data aug. method & Accuracy (CNN) & Accuracy (CRNN) \\
  \midrule 
  US8K-PS & 66.49 (0.60) & 65.01 (0.95)\\ 
  US8K-TS & 64.14 (0.80) & 62.63 (1.80) \\
  US8K-AudioGen & 68.42 (0.71) & 65.18 (0.87) \\
  US8K-AudioGen\textsubscript{gpt} & 68.88 (0.50) & \textbf{65.39 (0.63)} \\
  US8K-AudioLDM2 & 68.04 (0.63) & 63.41 (0.99) \\
  US8K-AudioLDM2\textsubscript{gpt} & \textbf{69.64 (0.91)} & 64.69 (0.53) \\
  \midrule
  US8K (Baseline) & 64.68 (0.82) & 62.70 (0.65)  \\ 
  \bottomrule
  \end{tabular}
 \label{tab:da}
\end{table} 
\subsection{Model Architectures}
\label{subsec:models}
We purposely select two simple, yet still relevant, architectures for ESC classification since our objective is to focus as much as possible on the quality of the data and not on the complexity of the architectures. Specifically, we considered a Convolutional Neural Network (CNN) and a Convolutional Recurrent Neural Network (CRNN) as ESC models. 
The CNN is implemented following a similar structure as the one presented in~\cite{salamon2017deep}. It is composed of three convolutional layers, each followed by a max-pooling operation, except the last layer. The kernel size, max pooling operation, and dropout parameters are the same as~\cite{salamon2017deep}. The CRNN is inspired by~\cite{castorena2023safety}. It is composed of seven convolutional blocks followed by a bidirectional GRU layer and a dense layer that generates the final output. We used the same parameters and configuration proposed in~\cite{castorena2023safety}. For consistency, the input of both networks consists of TF patches of $3~\mathrm{s}$ taken from the log mel-spectrogram computed from the audio input, as in~\cite{salamon2017deep}. All the sounds of US8K have been resampled to $16~\mathrm{kHz}$, being this the frequency at which the selected TTA models generate sounds. We computed the STFT considering a Hann window of $1024$ samples, and $2048$ frequency points. We used $64$ mel-bands for the log mel-spectrogram with a frequency range between $0~\mathrm{Hz}$ and $8000~\mathrm{Hz}$. Both networks have been trained for $100$ epochs, with batch size of $128$ and an early stop condition with patience on the validation loss of $15$ epochs. We considered Adam optimizer %~\cite{KingBa15}
with a learning rate of $0.001$. Samples shorter than $4~\mathrm{s}$ have been padded by repeating the sample until reaching the desired time length.
Our implementation of the networks is slightly different than the originals so, as is common in practice, results will not be exactly the same as the original paper.

\begin{figure}[t!]
\centering
\begin{subfigure}[b]{.9\columnwidth}
\centering
\includegraphics[width=\columnwidth]{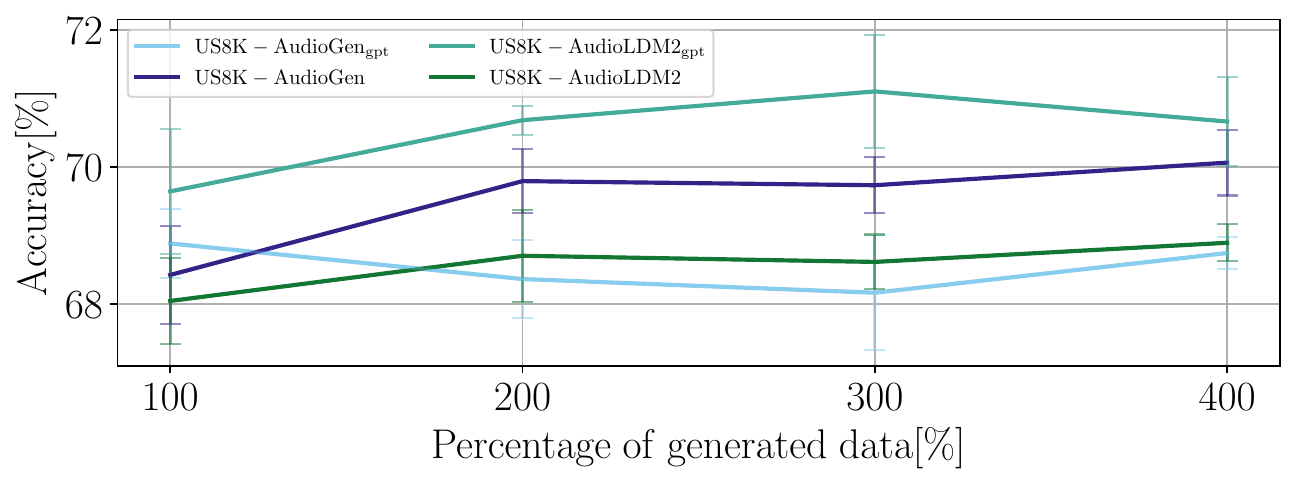}
 \caption{CNN}
 \label{subfig:augment_cnn}
\end{subfigure}
\hfill
\begin{subfigure}[b]{.9\columnwidth}
 \centering
\includegraphics[width=\columnwidth]{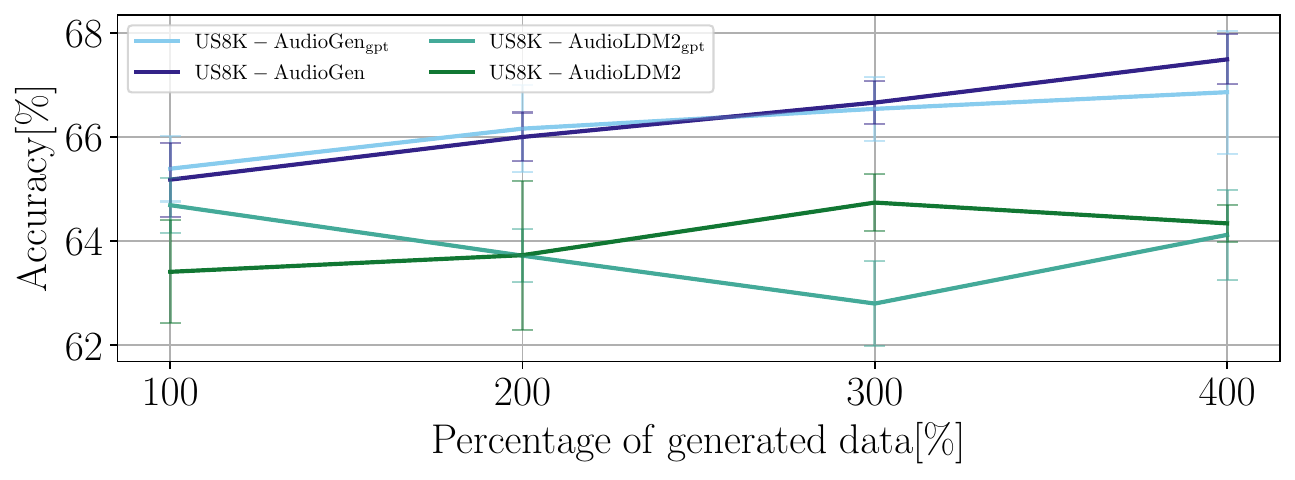}
 \caption{CRNN}
 \label{subfig:augment_crnn}
\end{subfigure}
\caption{Classification accuracy when varying the size of the TTA-generated augmentation dataset. Error bars represent 95\% confidence intervals over $5$ runs of the experiment.}
\label{fig:accuracy_vs_training_size}
\end{figure}

\section{Experiments and results}
\label{subse:expdef}
This section describes the experiments designed to understand to what extent the TTA-generated dataset impacts the performance of ESC learning-based models.
All results are averaged over 5 different runs of the whole 9-fold cross-validation. When referring to the baseline, we mean the ESC models trained with the original version of the US8K dataset.

\subsection{Can TTA-augmented datasets increase the accuracy of ESC models?}
\label{subsec:exp1}

This first experiment aims to understand how the integration of TTA-generated audio samples as a data augmentation technique affects the accuracy of the considered ESC systems. Data augmentation is a common technique used in ESC tasks to increase and diversify the dataset and improve the performance. Different techniques have been proposed in previous years~\cite{salamon2017deep, abayomi2022data} using signal processing-based methods.

To investigate this, we trained the two networks with the original US8K dataset by augmenting it with one version of the TTA-generated datasets.
We also compared the results with two signal processing data augmentation techniques: Time Stretching (TS), which is the process of changing the speed of an audio signal without affecting its pitch, and Pitch Shifting (PS), which is the process of changing the pitch without affecting the speed of the audio sample. While it would have been possible to compare with several augmentation techniques, we chose TS and PS since they are well established in the literature~\cite{abayomi2022data}.
For this study, we considered the same range of values of PS1 and TS in~\cite{salamon2017deep}. In all experiments, for each file, we randomly select only one PS and TS value between the four proposed in~\cite{salamon2017deep} to double the USK8 size. Table~\ref{tab:da} reports the accuracy results for the data augmentation techniques considered. \textit{USK-PS} and \textit{USK-TS} stand for PS and TS applied to the US8K dataset, respectively. The last row indicates the accuracy of the ESC systems trained with only the original US8K dataset. The results show that almost all the TTA-based augmentation techniques reach higher performances compared to the signal processing ones. For both models, the best accuracy scores are reached when GPT-based datasets are considered as data augmentation. The CNN model yields its optimal performance when augmented with \textit{AudioLDM2\textsubscript{gpt}}, achieving nearly a $5\%$ increase in accuracy over the baseline. For the CRNN model, the best results are obtained using \textit{AudioGen\textsubscript{gpt}}, reaching a $3\%$ enhancement compared to the baseline. 
Signal processing data augmentation techniques consistently yield inferior or comparable performances. These findings suggest that incorporating TTA-generated audio samples as a data augmentation technique enhances the performance of the ESC system.

Motivated by these results, we perform a further experiment to understand if increasing the size of the TTA-generated dataset leads to a corresponding increase in performance. We consecutively double the size of the data used for augmentation, up to $400\%$ the original size. We increased the size of the dataset following the same distribution of US8K. Results are reported in Fig.~\ref{fig:accuracy_vs_training_size}, where 100\% corresponds to the previous experiment. 

Although the CRNN shows improved performance when the original dataset is augmented by 200\% to 300\% with data from one of the two AudioGen-generated versions, no clear trend is observed for either model. 
%These results highlight the need for further investigation to understand the implications of the observed patterns.
These results suggest that using TTA models for data augmentation is not trivial and requires further investigation. 

\subsection{Can we rely on only TTA-generated data to train an ESC system?}
\label{subsec:exp2}
\begin{table}[t!]
\footnotesize
\caption{Models accuracy when trained only with synthetic data.}
\centering
 \begin{tabular}{l|c|c}
  \toprule
  Training dataset & Accuracy (CNN) & Accuracy (CRNN) \\
  \midrule 
  AudioGen & 40.32 (0.29) & 38.79 (1.24) \\
  AudioGen\textsubscript{gpt} & \textbf{46.04 (0.71)} & \textbf{43.96 (1.36)}\\
  AudioLDM2 & 38.81 (0.56) & 36.11 (1.11) \\
  AudioLDM2\textsubscript{gpt} & 38.49 (1.21) & 32.86 (1.01)\\
  \midrule
  US8K (Baseline) & 64.68 (0.82) & 62.70 (0.65)  \\ 
  \bottomrule
  \end{tabular}
 \label{tab:tta}
\end{table} 

Motivated by previous results, we explore if TTA-generated data alone can effectively train an ESC system. We trained the ESC models with the different TTA-generated versions of the dataset and tested the models on real data. Table~\ref{tab:tta} reports the accuracy for the different cases compared with the baseline. The baseline achieves the best performance. However, it is worth noticing that both ESC models (when trained with synthetic data) achieve their highest accuracy when using AudioGen\textsubscript{gpt} dataset. This emphasizes the preference for AudioGen as a TTA model as a dataset generator for ESC. In contrast, using AudioLDM2 results in inferior performance. However, the results suggest that depending solely on TTA-generated datasets is not yet feasible. Our intuition is that domain adaptation between the TTA-generated used for training and real data used for testing impacts the performances and this will be explored in future investigations. 

\begin{figure}[t!]
\centering
\begin{subfigure}[b]{.9\columnwidth}
\centering
\includegraphics[width=\columnwidth]{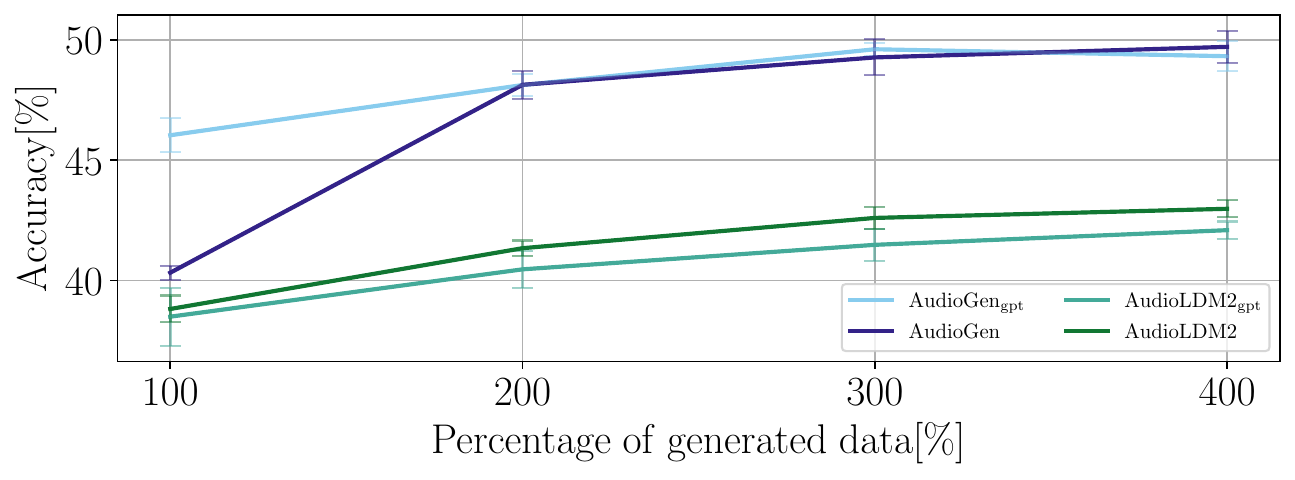}
 \caption{CNN}
 \label{subfig:sim_only_cnn}
\end{subfigure}
\vfill
\begin{subfigure}[b]{.9\columnwidth}
 \centering
\includegraphics[width=\columnwidth]{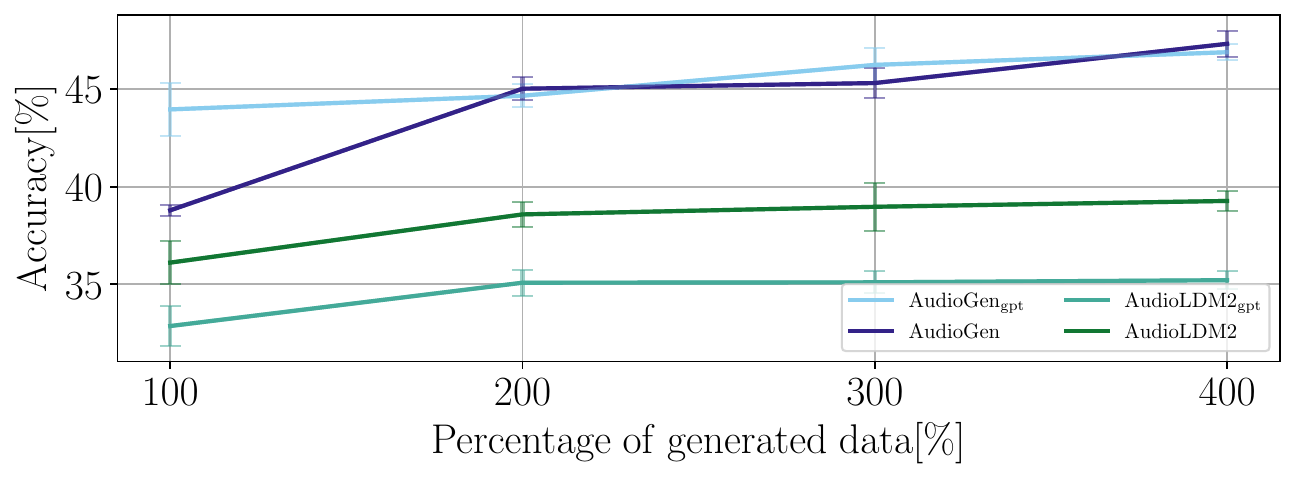}
 \caption{CRNN}
 \label{subfig:sim_only_crnn}
\end{subfigure}
\caption{Classification accuracy when varying the size of the training dataset composed of only TTA-generated data. Error bars: 95\% confidence intervals over $5$ experiment repetitions.}
\label{fig:accuracy_vs_training_size_gen_data_only}
\end{figure}
 
As for the previous experiment, we analyzed if the threshold for achieving baseline performance might be influenced by the quantity of data used at training. Also in this case, we incrementally doubled the dataset size to train models with up to 400\% of synthetic data. As reported in Fig.~\ref{fig:accuracy_vs_training_size_gen_data_only}, increasing the number of audio data is useful up to 2-3 times the original dataset size, confirming the previous case experiment. Also in this case, both networks achieve higher performances when trained with AudioGen dataset versions, suggesting that AudioGen has the capabilities of generating more realistic audio data that can be used to train ESC models that will then be able to better generalize to real-world data.
%probably due to its capabilities in generating more realistic sounds 
%\vspace{-3em}

\subsection{To what extent real data can be safely replaced by synthetic data generated through TTA models?}
\label{subsec:exp3}
Previous findings showed that datasets generated through TTA models can enhance performance when used for data augmentation, but solely using synthetic data is not sufficient for effectively training an ESC model. These observations make us wonder if there is a threshold at which synthetic data can effectively replace real data, allowing an ESC model to achieve baseline or better performance while requiring less real data. To investigate this, we conducted several
experiments where we incrementally replaced one or more folders of the real US8K dataset with corresponding TTA-generated synthetic folders. Starting with replacing one folder and progressing up to eight folders, we ensured that at least one folder of real data was always included in the training dataset. The folder to be replaced was randomly selected for each iteration of the single experiment. Results are reported in Fig.~\ref{fig:mixed_dataset}. The straight line indicates the baseline performance. Both ESC models have a similar trend: with up to nearly 20\% of real data replaced by synthetic data, the performance is comparable and slightly better for the CRNN. However, beyond this point and up to almost 50\% replacement, the 
accuracy begins to decrease, losing nearly 10\%. A noticeable drop in performance occurs beyond the 50\% replacement level, with the decline becoming steeper as more real audio files are replaced, ultimately reaching a performance level similar to the experiment described in Section~\ref{subsec:exp2} when 8 out of 9 folders are replaced. It is worth noting that the AudioGen\textsubscript{gpt} version of the dataset allows the model to maintain comparable performance even when about 40\% of the data is synthetically generated.

\begin{figure}[t!]
\centering
\begin{subfigure}[b]{.9\columnwidth}
\centering
\includegraphics[width=\columnwidth]{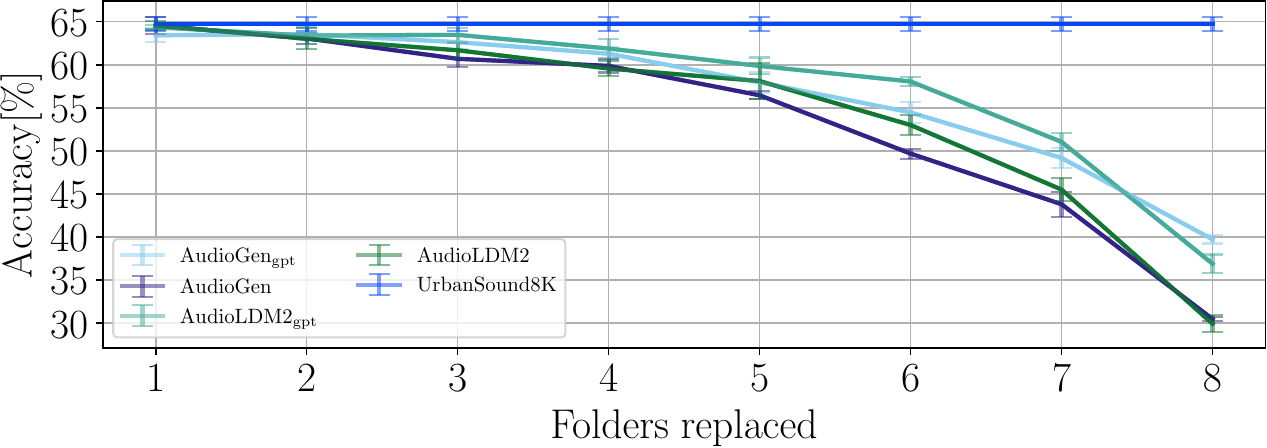}
 \caption{CNN}
 \label{subfig:folds_sim_cnn}
\end{subfigure}
\hfill
\begin{subfigure}[b]{.9\columnwidth}
 \centering
\includegraphics[width=\columnwidth]{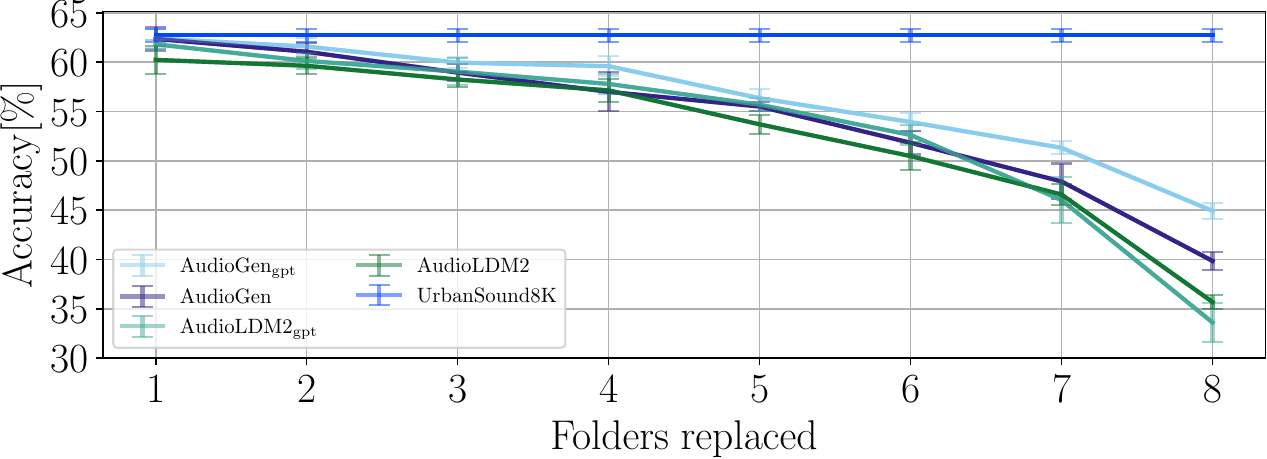}
 \caption{CRNN}
 \label{subfig:folds_sim_crnn}
\end{subfigure}
\caption{Classification accuracy when incrementally replacing US8K folders using TTA-generated data. Error bars: 95\% confidence intervals over 5 experiment repetitions.}
\label{fig:mixed_dataset}
\end{figure}

\section{Discussion}
The results show that when training ESC models, TTA-generated data are useful when used to augment or replace part of the real dataset, but they are not ready to completely replace it. While a complete analysis of the reason behind this is out of the scope of this paper and would require further investigations, we report here a few anecdotal causes that we encountered. TTA models do not generate audio related to all the classes with the same effectiveness. For example, a dog barking is better reproduced compared to an audio clip of street music; hammer and air conditioning sounds might be too similar, etc. This is probably part of the reason why the performance drastically drops when only generated data are used during training. We also conducted a preliminary experiment by removing the street music class from the dataset (both training and evaluation), which is the most problematic class.
However, no better results were obtained. 

Interestingly, the results of this study are in line with the outcome of a parallel study that came out at the time of writing~\cite{feng2024can}, where it is reported a consistent drop in performances when using only synthetic data for similar tasks. %and it is shown how LLM-based prompts are preferable to enrich the quality of the data generated for training. 

\section{Conclusions and future works}
\label{sec:conclusions}

This paper investigates the impact of incorporating Text-To-Audio-generated datasets into the training process of ESC systems. We conducted various experiments to explore different methods of integrating and replacing the original dataset with additional training data generated with TTA models. The results show that generated datasets are beneficial when used as data augmentation techniques, but are not ready to be used as the only source of data during training. When replacing part of the real dataset with synthetically generated data, the results are comparable to the baseline up to 10-20\% of the data, depending on the ESC model and TTA used. 
We believe that the obtained results motivate further investigations on the topic. In fact, as the quality of TTAs increases, it is likely that such a training set synthesis approach will be more and more beneficial. Future works will include the exploration of more advanced prompt engineering strategies and the investigation of fine-tuning methods to improve the generation capabilities of TTA models.

\bibliographystyle{IEEEtran}
\bibliography{refs}

\begin{thebibliography}{10}
\providecommand{\url}[1]{#1}
\def\UrlFont{\rmfamily}
\providecommand{\newblock}{\relax}
\providecommand{\bibinfo}[2]{#2}
\providecommand\BIBentrySTDinterwordspacing{\spaceskip=0pt\relax}
\providecommand\BIBentryALTinterwordstretchfactor{4}
\providecommand\BIBentryALTinterwordspacing{\spaceskip=\fontdimen2\font plus
\BIBentryALTinterwordstretchfactor\fontdimen3\font minus \fontdimen4\font\relax}
\providecommand\BIBforeignlanguage[2]{{%
\expandafter\ifx\csname l@#1\endcsname\relax
\typeout{** WARNING: IEEEtran.bst: No hyphenation pattern has been}%
\typeout{** loaded for the language `#1'. Using the pattern for}%
\typeout{** the default language instead.}%
\else
\language=\csname l@#1\endcsname
\fi
#2}}

\bibitem{kreuk2022audiogen}
F.~Kreuk, G.~Synnaeve, A.~Polyak, U.~Singer, A.~D{\'e}fossez, J.~Copet, D.~Parikh, Y.~Taigman, and Y.~Adi, ``Audiogen: Textually guided audio generation,'' in \emph{ICLR}, 2022.

\bibitem{liu2023audioldm}
H.~Liu, Z.~Chen, Y.~Yuan, X.~Mei, X.~Liu, D.~Mandic, W.~Wang, and M.~D. Plumbley, ``Audioldm: Text-to-audio generation with latent diffusion models,'' \emph{arXiv:2301.12503}, 2023.

\bibitem{rombach2022high}
R.~Rombach, A.~Blattmann, D.~Lorenz, P.~Esser, and B.~Ommer, ``High-resolution image synthesis with latent diffusion models,'' in \emph{Proc. IEEE/CVF CVPR}, 2022.

\bibitem{liu2023audioldm2}
H.~Liu, Q.~Tian, Y.~Yuan, X.~Liu, X.~Mei, Q.~Kong, Y.~Wang, W.~Wang, Y.~Wang, and M.~D. Plumbley, ``Audioldm 2: Learning holistic audio generation with self-supervised pretraining,'' \emph{arXiv:2308.05734}, 2023.

\bibitem{ghosal2023text}
D.~Ghosal, N.~Majumder, A.~Mehrish, and S.~Poria, ``Text-to-audio generation using instruction-tuned llm and latent diffusion model,'' \emph{arXiv:2304.13731}, 2023.

\bibitem{huang2023make}
R.~Huang, J.~Huang, D.~Yang, Y.~Ren, L.~Liu, M.~Li, Z.~Ye, J.~Liu, X.~Yin, and Z.~Zhao, ``Make-an-audio: Text-to-audio generation with prompt-enhanced diffusion models,'' in \emph{ICML}.\hskip 1em plus 0.5em minus 0.4em\relax PMLR, 2023.

\bibitem{vyas2023audiobox}
A.~Vyas, B.~Shi, M.~Le, A.~Tjandra, Y.-C. Wu, B.~Guo, J.~Zhang, X.~Zhang, R.~Adkins, W.~Ngan, \emph{et~al.}, ``Audiobox: Unified audio generation with natural language prompts,'' \emph{arXiv:2312.15821}, 2023.

\bibitem{nordahl2011sound}
R.~Nordahl, L.~Turchet, and S.~Serafin, ``Sound synthesis and evaluation of interactive footsteps and environmental sounds rendering for virtual reality applications,'' \emph{Trans. Vis. Comput. Graph.}, 2011.

\bibitem{chung2024t}
Y.~Chung, J.~Lee, and J.~Nam, ``T-foley: A controllable waveform-domain diffusion model for temporal-event-guided foley sound synthesis,'' \emph{arXiv:2401.09294}, 2024.

\bibitem{ronchini2021impact}
F.~Ronchini, R.~Serizel, N.~Turpault, and S.~Cornell, ``The impact of non-target events in synthetic soundscapes for sound event detection,'' in \emph{DCASE Workshop}, 2021.

\bibitem{salamon2017scaper}
J.~Salamon, D.~MacConnell, M.~Cartwright, P.~Li, and J.~P. Bello, ``Scaper: A library for soundscape synthesis and augmentation,'' in \emph{Proc. WASPAA}.\hskip 1em plus 0.5em minus 0.4em\relax IEEE, 2017.

\bibitem{ronchini2022benchmark}
F.~Ronchini and R.~Serizel, ``A benchmark of state-of-the-art sound event detection systems evaluated on synthetic soundscapes,'' in \emph{Proc. ICASSP}.\hskip 1em plus 0.5em minus 0.4em\relax IEEE, 2022.

\bibitem{gontier2021polyphonic}
F.~Gontier, V.~Lostanlen, M.~Lagrange, N.~Fortin, C.~Lavandier, and J.-F. Petiot, ``Polyphonic training set synthesis improves self-supervised urban sound classification,'' \emph{The Journal of the Acoustical Society of America}, 2021.

\bibitem{damiano}
S.~Damiano, L.~Bondi, S.~Ghaffarzadegan, A.~Guntoro, and T.~van Waterschoot, ``Can synthetic data boost the training of deep acoustic vehicle counting networks?'' in \emph{Proc. ICASSP}, 2024.

\bibitem{ibrahim2024towards}
K.~M. Ibrahim, A.~Perzo, and S.~Leglaive, ``Towards improving speech emotion recognition using synthetic data augmentation from emotion conversion,'' in \emph{Proc. ICASSP}, 2024.

\bibitem{damiano2022pyroadacoustics}
S.~Damiano and T.~van Waterschoot, ``Pyroadacoustics: a road acoustics simulator based on variable length delay lines,'' in \emph{Proc. 25th Int. Conf. Digital Audio Effects}, 2022.

\bibitem{kroher2023can}
N.~Kroher, H.~Cuesta, and A.~Pikrakis, ``Can musicgen create training data for mir tasks?'' \emph{arXiv:2311.09094}, 2023.

\bibitem{copet2024simple}
J.~Copet, F.~Kreuk, I.~Gat, T.~Remez, D.~Kant, G.~Synnaeve, Y.~Adi, and A.~D{\'e}fossez, ``Simple and controllable music generation,'' \emph{NeurIPS}, 2024.

\bibitem{zhang2023first}
H.~Zhang, Q.~Zhu, J.~Guan, H.~Liu, F.~Xiao, J.~Tian, X.~Mei, X.~Liu, and W.~Wang, ``First-shot unsupervised anomalous sound detection with unknown anomalies estimated by metadata-assisted audio generation,'' \emph{arXiv:2310.14173}, 2023.

\bibitem{bansal2022environmental}
A.~Bansal and N.~K. Garg, ``Environmental sound classification: A descriptive review of the literature,'' \emph{ISWA}, 2022.

\bibitem{feng2024can}
T.~Feng, D.~Dimitriadis, and S.~Narayanan, ``Can synthetic audio from generative foundation models assist audio recognition and speech modeling?'' \emph{arXiv e-prints}, 2024.

\bibitem{wu2023large}
Y.~Wu, K.~Chen, T.~Zhang, Y.~Hui, T.~Berg-Kirkpatrick, and S.~Dubnov, ``Large-scale contrastive language-audio pretraining with feature fusion and keyword-to-caption augmentation,'' in \emph{Proc. ICASSP}.\hskip 1em plus 0.5em minus 0.4em\relax IEEE, 2023.

\bibitem{salamon2014dataset}
J.~Salamon, C.~Jacoby, and J.~P. Bello, ``A dataset and taxonomy for urban sound research,'' in \emph{Proc. ACM Multimed.}, 2014.

\bibitem{piczak2015esc}
K.~J. Piczak, ``{ESC}: Dataset for environmental sound classification,'' in \emph{Proc. ACM Multimed.}, 2015.

\bibitem{elizalde2023clap}
B.~Elizalde, S.~Deshmukh, M.~Al~Ismail, and H.~Wang, ``Clap learning audio concepts from natural language supervision,'' in \emph{Proc. ICASSP}.\hskip 1em plus 0.5em minus 0.4em\relax IEEE, 2023.

\bibitem{deshmukh2024pam}
S.~Deshmukh, D.~Alharthi, B.~Elizalde, H.~Gamper, M.~A. Ismail, R.~Singh, B.~Raj, and H.~Wang, ``Pam: Prompting audio-language models for audio quality assessment,'' \emph{arXiv preprint arXiv:2402.00282}, 2024.

\bibitem{he2022synthetic}
R.~He, S.~Sun, X.~Yu, C.~Xue, W.~Zhang, P.~Torr, S.~Bai, and X.~Qi, ``Is synthetic data from generative models ready for image recognition?'' \emph{arXiv preprint arXiv:2210.07574}, 2022.

\bibitem{mei2023wavcaps}
X.~Mei, C.~Meng, H.~Liu, Q.~Kong, T.~Ko, C.~Zhao, M.~D. Plumbley, Y.~Zou, and W.~Wang, ``Wavcaps: A chatgpt-assisted weakly-labelled audio captioning dataset for audio-language multimodal research,'' \emph{IEEE/ACM Trans. Audio Speech Lang. Process.}, 2024.

\bibitem{voetman2023using}
R.~Voetman, A.~van Meekeren, M.~Aghaei, and K.~Dijkstra, ``Using diffusion models for dataset generation: Prompt engineering vs. fine-tuning,'' in \emph{Proc. CAIP}.\hskip 1em plus 0.5em minus 0.4em\relax Springer, 2023.

\bibitem{deshmukh2024domain}
S.~Deshmukh, R.~Singh, and B.~Raj, ``Domain adaptation for contrastive audio-language models,'' \emph{arXiv preprint arXiv:2402.09585}, 2024.

\bibitem{salamon2017deep}
J.~Salamon and J.~P. Bello, ``Deep convolutional neural networks and data augmentation for environmental sound classification,'' \emph{IEEE Signal Process. Lett.}, 2017.

\bibitem{castorena2023safety}
C.~Castorena, M.~Cobos, J.~Lopez-Ballester, and F.~J. Ferri, ``A safety-oriented framework for sound event detection in driving scenarios,'' \emph{Applied Acoustics}, 2023.

\bibitem{abayomi2022data}
O.~O. Abayomi-Alli, R.~Dama{\v{s}}evi{\v{c}}ius, A.~Qazi, M.~Adedoyin-Olowe, and S.~Misra, ``Data augmentation and deep learning methods in sound classification: A systematic review,'' \emph{Electronics}, 2022.

\end{thebibliography}

\end{sloppy}
\end{document}